\documentclass[fleqn,usenatbib]{mnras}

\usepackage[T1]{fontenc}
\usepackage{ae,aecompl}

\usepackage{graphicx}	
\usepackage{amsmath, bm}	
\usepackage{amssymb}	
\usepackage{lipsum}

\newcommand{\kmpers}{\,\rm{\,km \, s^{-1}}} 
\newcommand{\kpc}{\,\rm{kpc}} 
\newcommand{\masslossrate}{\,\rm{M_{\odot} yr^{-1}}} 

\title[Probabilistic orbits]{Probabilistic orbits and dynamical masses of emission-line binaries}

\author[D. Grant and K. Blundell]{
David Grant$^{1}$\thanks{E-mail: david.grant@physics.ox.ac.uk}
and Katherine Blundell$^{1}$
\\
$^{1}$University of Oxford, Department of Physics, Keble Road, Oxford, OX1 3RH, U.K\\
}

\date{Accepted xxxx. Received yyyy; in original form zzzz}

\pubyear{2021}

\begin{document}
\label{firstpage}
\pagerange{\pageref{firstpage}--\pageref{lastpage}}
\maketitle

\begin{abstract}
The observed orbits of emission-line stars may be affected by systematics owing to their broad emission lines being formed in complex and extended environments. This is problematic when orbital parameter probability distributions are estimated assuming radial-velocity data are solely comprised of Keplerian motion plus Gaussian white noise, leading to overconfident and inaccurate orbital solutions, with implications for the inferred dynamical masses and hence evolutionary models. We present a framework in which these systems can be meaningfully analysed. We synthesise benchmark datasets, each with a different and challenging noise formulation, for testing the performance of different algorithms. We make these datasets freely available with the aim of making model validation an easy and standardised practice in this field. Next, we develop an application of Gaussian processes to model the radial-velocity systematics of emission-line binaries, named marginalised $\mathcal{GP}$. We benchmark this algorithm, along with current standardised algorithms, on the synthetic datasets and find our marginalised $\mathcal{GP}$ algorithm performs significantly better than the standard algorithms for data contaminated by systematics. Finally, we apply the marginalised $\mathcal{GP}$ algorithm to four prototypical emission-line binaries: Eta Carinae, GG Carinae, WR 140, and WR 133. We find systematics to be present in several of these case studies; and consequently, the predicted orbital parameter distributions, and dynamical masses, are modified from those previously determined.
\end{abstract}

\begin{keywords}
methods: statistical -- binaries: spectroscopic -- binaries: visual -- stars: emission-line, Be -- stars: individual: Eta Carinae, GG Carinae, WR, 140, WR 133
\end{keywords}



\section{Introduction}
\label{sec:introduction}
Accurate inferences of the orbits and masses of multiple-star systems are required in order to fully understand the lives and deaths of stars. For binary systems, well-sampled time-series spectroscopic measurements enable the line-of-sight projected orbital motion to be calculated. When these radial-velocity measurements are combined with either interferometric imaging -- or light curves in the case of eclipsing binaries -- the full three-dimensional orbit can be constrained \citep[eg.][]{Thomas2021The140, Richardson2021TheOrbit}. As a result, accurate dynamical masses of the individual stellar components can be computed. However, these results are predicated on the assumption that the radial-velocity data well-represents the orbital motion of the system. Whilst this is a fair assumption for many stars, where the spectral lines are formed unencumbered at the stellar surface, for emission-line stars the spectral lines may encode velocities that are not purely orbital in their origin \citep[][]{Grant2020Uncovering140}. Consequently, radial-velocity data for emission-line binaries may often be afflicted by systematic errors with implications for the inferred orbits and dynamical masses.

The sources of possible systematics are numerous. The broad emission lines are formed in a complex environment of outflowing gas and often in the presence of colliding-wind excess emission, both of which can alter the line profiles and bias the measurements. Additionally, methods to extract the radial velocities may be imperfect at disentangling the portion of the line profiles that represents solely the Keplerian motion.

The current standard practice for emission-line binaries is to assume radial velocities are ideally extracted and exhibit Keplerian motion plus some Gaussian white noise. Orbital models are then fitted to the data and the parameters of the orbit are estimated. Any poorly fit data are often noted qualitatively and thereafter ignored as the orbital parameters are carried through to the subsequent analysis of the stellar masses and evolutionary state of the systems. Put bluntly: if the orbital models do not explain all of the signals in the data then we are at risk of inferring biased results.

In this study we want to create a sea change in the methodology when estimating the orbital parameters of emission-line binaries. In Section \ref{sec:benchmark_dataset} we synthesise benchmark datasets, with various noise formulations, for testing the robustness of orbital models in single-lined spectroscopic binaries. We make these datasets freely available with the aim of making model validation an easy and standardised practice. In Section \ref{sec:benchmark_algorithm} we describe the current algorithms used within the field, and introduce a new approach to modelling the radial velocities of emission-line binaries using Gaussian processes (GPs). We apply the algorithms to the benchmark datasets and find the GP algorithm serves as the current best method. We encourage other researchers to employ novel techniques to improve upon our scores with these same synthetic datasets. In Section \ref{sec:discussion} we discuss how the results can be affected by having different data available, such as in the case of a double-lined spectroscopic binary or by having access to relative astrometric measurements. We then apply our GP algorithm to four real systems -- Eta Carinae, GG Caraine, WR 140, and WR 133 -- to assess the impact of our new approach on the estimated orbital parameters and dynamical masses. Finally, in Section \ref{sec:summary_and_conclusions} we summarise our findings.

\section{A benchmark dataset}
\label{sec:benchmark_dataset}
We conceive of a data analysis work flow in which the investigator -- analysing radial-velocity data in our use case -- first tests their models on a benchmark dataset, before fitting their models to their observations. In this way the investigator can ensure their model is robust and performant, as well as providing a quantitative measure by which the best modelling techniques can take precedence in the field. This methodology is inspired by other fields, such as machine learning, in which advancements have long been validated using benchmark datasets.

To this end, we make a first attempt at creating benchmark datasets for the orbits of emission-line binaries. We restrict the data to single-lined spectroscopic binaries (SB1), however we discuss more complex data combinations in Section \ref{sec:discussion}. We generate three datasets each having the same number of synthetic binaries, the same distribution of orbital parameters, but with different noise formulations. 

The orbital parameters required to fully specify SB1 data are the time of periastron, $T_0$, the orbital period, $P$, the eccentricity, $e$, the argument of periastron, $\omega$, the primary star's semi-amplitude, $k_1$, and the primary star's radial-velocity offset, $\gamma_1$. We fix $T_0=2458932$ (JD), $P=50$ days, and $\gamma_1=0 \kmpers$ as these parameters only serve to re-scale or translate the data in the time or velocity dimensions. We select 100 observations from one orbital period for every system. The observations are drawn from a beta distribution with shape parameters $\alpha = \beta = 1 - e$ with values corresponding to the orbital phase. The result is a realistic imitation of observing patterns conducted on binary stars, in which observations around periastron become more important as the eccentricity increases, and there is always sufficient orbital coverage to infer all six of the orbital parameters.

The noise added to each dataset is designed to challenge potential models to be robust against the various problems that may be embedded in the radial velocities extracted from emission-line stars. In dataset 1 we add only Gaussian white noise to the velocities. This is the easiest dataset for models to contend with, resembling the case of ideally extracted observations encoding purely orbital motion. The white noise level is fixed to $5 \kmpers{}$ and so the value given to $k_1$ serves to adjust the signal-to-noise level. In dataset 2 we add correlated noise to the velocities by injecting randomly generated polynomials. The polynomials are synthesised from a random set of roots that lie within the observational time frame. For each system, the polynomials are adjusted in amplitude by scaling the maximum deviation by a sample from the uniform distribution $0 - 0.5k_1$, and adjusted in their time-variability by sampling an integer number of roots uniformly over the range 1-10. In dataset 3 we again add correlated noise to the velocities, but in this case we inject functions randomly drawn from a Gaussian process prior. For each system, we tune the amplitude and time-variability of the interloping signal through the squared exponential covariance function. We use this kernel as it produces smoothly-varying functions in time. For each system we sample an amplitude scale factor from the uniform distribution $0 - 0.5k_1$, and adjust the time-variability by sampling the characteristic length scale from the uniform distribution $0.1 - 30$ days. A more thorough definition of a Gaussian process is treated in Section \ref{sec:benchmark_algorithm}.

The synthesised systems each have a total of five parameters that are varied across the orbital model and noise formulation. In order to balance the time required to benchmark a model across the datasets with good parameter coverage, we generate $N=3^5=243$ systems for each of the three datasets\footnote{Our models run for approximately 17 minutes per synthetic object on one 2 GHz Intel core i5 processor, or eg. 3.4 hours for an entire dataset on 20 cores assuming ideal parallelisation.}. For each system the data included are the radial-velocity measurements, $\bm{y} = (y_1,...,y_N)^T$, each at a given time and spectral line energy, $\bm{X} = (\bm{x}_1,...,\bm{x}_N)^T = ((t_1, E_1)^T,...,(t_N, E_N)^T)^T$, and with associated uncertainties, measuring the white noise only, $\bm{\sigma} = (\sigma_1,...,\sigma_N)^T$, as well as the true orbital parameters.

The benchmark datasets are open-source and freely available\footnote{\url{https://github.com/DavoGrant/}}. The data are provided in a number of easily usable formats and we encourage researchers to validate their current models with the data, and to beat the scores we set in this study with new and ingenious models.

\begin{figure*}
	\includegraphics[width=\textwidth]{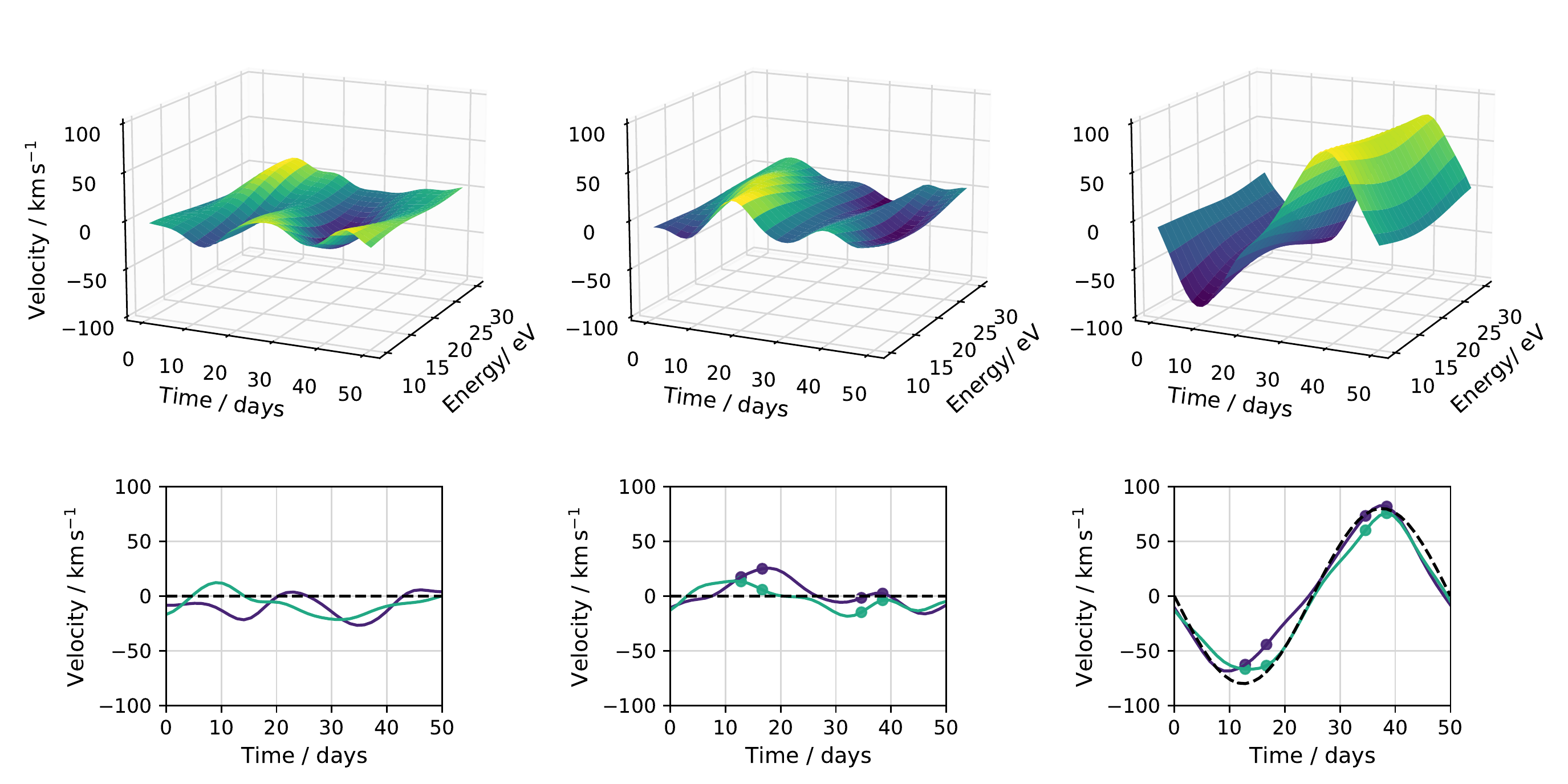}
    \caption{A visual representation of the GP algorithm. The top row displays the model spanning the full two-dimensional grid, whilst the bottom row shows two slices at energies 15 eV (purple line) and 25 eV (green line). The left-hand column shows a sample taken from the GP prior. The middle column shows a sample taken from the GP conditioned on a small amount of input data with no uncertainty. The right-hand column shows the same GP as in the middle column but with the mean model added back in. The GP hyperparameters were fixed at $A=11^2 \kmpers$, $l_1=5^2$ days, $l_2=8^2$ eV, and $\sigma_w=0 \kmpers$ days.}
    \label{fig:gp_prior_condtional_and_mean}
\end{figure*}

\subsection{Evaluation criterion}
\label{subsec:evaluation_criterion}
In order to quantify the performance of different models on each benchmark dataset we define an evaluation criterion. The criterion will assess how well the inferences made about the orbital parameters compare to the injected true values. Following similar motivation as \citet[][]{Quinonero-Candela2005LNAIChallenge}, the desired behaviour of the criterion is such that for two models that estimate parameter distributions centred on the same values but with differing widths, the more confident model will score higher (lower) when the estimates are more accurate (inaccurate). The best scoring models will thus be those with accurate and confident parameter estimations, while inaccurate estimations will be penalised for overconfidence. To achieve this we compute the density of the predicted posterior distribution at the true value. For each of the assessable parameters this can be written as
\begin{equation}
p(\bm{w} = \bm{w}_{t} \mid \bm{y}, \bm{X}) = \frac{p(\bm{y} \mid \bm{X}, \bm{w}_{t}) p(\bm{w}_{t})}{p(\bm{y} \mid \bm{X})},
	\label{eq:evaluation_criterion}
\end{equation}
where $\bm{w}$ are the parameters of the model and $\bm{w}_t$ is the injected true parameter vector we wish to recover. The denominator, known as the marginal likelihood, is given by
\begin{equation}
p(\bm{y} \mid \bm{X}) =  \int p(\bm{y} \mid \bm{X}, \bm{w}) p(\bm{w}) d\bm{w}.
	\label{eq:evaluation_criterion_marginal_likelihood}
\end{equation}
The overall score per model per dataset is
\begin{equation}
\mathrm{score} =  \mathrm{median}(S),
	\label{eq:evaluation_criterion_total_score}
\end{equation}
where the set, $S$, is
\begin{equation}
S =  \{ \ln p(\bm{w} = \bm{w}_{t, i} \mid \bm{y}_i, \bm{X}_i) \mid i = 1, 2, \dots, N \},
	\label{eq:evaluation_criterion_total_score_verbose}
\end{equation}
and $N$ is the number of synthetic objects in the dataset. This score is the median logarithmic density of the predicted posterior distribution at the true value for each of the parameters. Note that the score can be negative and will depend on the scale of each parameter. In practice, sampling methods will be used to estimate the posterior distribution, owing to the difficulty in computing the marginal likelihood integral. We therefore employ kernel density estimation \citep[][]{Scott2015MultivariateEdition}, with Silverman's rule of thumb to select the bandwidth \citep[][]{Silverman2018DensityAnalysis}, to evaluate the score in a non-parametric way.

\section{A benchmark algorithm}
\label{sec:benchmark_algorithm}
In this section we introduce benchmark algorithms for analysing the orbits of emission-line binaries through Bayesian parameter estimation. We briefly re-state two common procedures used in the literature, as well as a generative algorithm based on Gaussian processes to account for systematic noise. These algorithms can be summarised as follows:
\begin{enumerate}
  \item $\chi^2$, this algorithm is the most simple in its procedure. It is based on the assumption that the radial-velocity data only encode the orbital motion plus some Gaussian white noise. In this case the objective function is proportional to the $\chi^2$ value.
  \item Marginalised $\sigma$, this algorithm is a simple extension to the first algorithm. It is used when the uncertainties output from radial-velocity extraction techniques do not appropriately represent the variance in the data about the orbital motion. An additional free parameter is added in quadrature to the extracted uncertainties in the data to account for the initially poorly estimated uncertainties. This term is sometimes referred to as a jitter in the literature and can be both statistical or systematic in origin.
  \item Marginalised $\mathcal{GP}$, this algorithm is comprised of a mean model plus a GP, to model the orbital motion plus any systematics respectively. The hyperparameters of the GP are marginalised over to best capture the predictive uncertainty in the orbital parameters.
\end{enumerate}

The third benchmark algorithm described above, marginalised $\mathcal{GP}$, takes a generative approach to dealing with the noise present in the radial-velocity data. A GP is defined as a collection of random variables, any finite number of which have a joint Gaussian distribution \citep[][]{Rasmussen2006GaussianLearning}. This non-parametric approach to regression lends itself to modelling the unknown noise component in astrophysics datasets. GPs have been used previously, for example, by \citet[][]{Brewer2009GaussianData} for stellar oscillations in light curves, by \citet[][]{Gibson2012ASpectroscopy} for transmission spectroscopy, by \citet[][]{Aksulu2020ASystematics} for gamma-ray burst afterglows, and by \citet[][]{Aigrain2016K2SC:Regression} for exoplanet light curves. Here we describe a novel application of GPs to the radial-velocity systematics of emission-line binaries.

\subsection{Mean models}
\label{subsec:mean_model}
At the heart of our marginalised $\mathcal{GP}$ algorithm we require a parameterised model of the orbiting emission-line stars. For radial velocities extracted from lines that form deep in the wind, the velocities can be described by Keplerian motion. However, \citet[][]{Grant2020Uncovering140} found that for lines emitted over extended radii, the observed motion is better characterised by a convolution of past motion accounting for the emission region of a particular spectral line. Consequently, we have two possible orbital models. For lines which form deep in the wind we have the \citet[][]{Aitken1964TheStars} standard expression for radial velocities
\begin{equation}
v_{\mathrm{kep}}(t) = k_1(\cos (\omega + \nu(t)) + e\cos{\omega}) + \gamma_1,
	\label{eq:pure_keplerian_mean_model}
\end{equation}
or, for lines forming over extended regions in the wind, we have the \citet[][]{Grant2020Uncovering140} modified formalism
\begin{equation}
v_{\mathrm{ckm},n}(t) = v_{\mathrm{kep}}(t) * \lambda_{L,n}(t) = \int_{-\infty}^{\infty} v_{\rm{kep}}(\tau) \lambda_{L,n}(t - \tau) d\tau,
	\label{eq:convolutional_keplerian_mean_model}
\end{equation}
where in these equations $\nu$ is the true anomaly, $v_{\rm{kep}}$ is the Keplerian velocity projected onto our line-of-sight, and $v_{\mathrm{ckm}, n}$ is the convolutional Keplerian motion which depends on the spectral line, $n$, via the line-formation kernel, $\lambda_{L,n}(t)$. 

For the generation and analysis of the benchmark datasets described in Section \ref{sec:benchmark_dataset} we exclusively use equation \ref{eq:pure_keplerian_mean_model} as the orbital model, which has the parameter vector
\begin{equation}
\bm{\theta} = (T_0, P, \omega, k_1, \gamma_1)^T.
	\label{eq:mean_model_parameter_vector}
\end{equation}
For some observational datasets analysed in Section \ref{subsec:application_to_prototypical} we also make use of equation \ref{eq:convolutional_keplerian_mean_model}. These models are the basis for all three benchmark algorithms, but specifically for the GP algorithm they are referred to as the mean model. 

\subsection{Gaussian processes}
\label{subsec:gaussina_process}
We wrap the mean model in a GP framework to model additional systematics; these could be noise or other non-orbital astrophysical signals. More formally this is written
\begin{equation}
f(\bm{X}) \sim \mathcal{GP} \big( v(\bm{X}, \bm{\theta}), k(\bm{X}, \bm{\phi}) \big),
	\label{eq:gaussina_process}
\end{equation}
where $v$ is the mean model, $k$ is the covariance function, $\bm{\theta}$ is the parameter vector of the mean model, and $\bm{\phi}$ is the hyperparameter vector of the covariance function.

Through specifying the covariance function we are encoding our assumptions about the nature of the systematics present in our data. We select the squared exponential kernel with the aim of capturing smoothly-varying deviations from orbital motion. For two input points, the covariance is described as
\begin{equation}
k(\bm{X}_p, \bm{X}_q) = A \exp \Bigg[-\frac{1}{2} \sum_{r=1}^2 \frac{(\bm{X}_{pr} - \bm{X}_{qr})^2}{l_r^2} \Bigg] + \delta_{pq} \sigma_w^2,
	\label{eq:covaraince_function_sqaured_exponential}
\end{equation}
where $A$ controls the amplitude, $l_r$ controls the characteristic length scale of the correlations in each input dimension, and $\sigma_w$ represents the white noise. From this expression it is clear that the closer input values are together, the more correlated the target values are expected to be, plus a certain level of white noise. The white noise parameter is further broken down into $\sigma_w^2 = \sigma_a^2 + \sigma_b^2$ where $\sigma_a^2$ is the extracted variance of the white noise and $\sigma_b^2$ is a free parameter which helps to account for underestimated values of $\sigma_a^2$.

We formulate the covariance in two dimensions to capture systematics that are correlated in both time and spectral-line energy. This follows from previous work on the dynamics of emission-line stars by \citet[][]{Grant2020Uncovering140} and \citet[][]{Porter2021GGPhotometry} in which the deviations from orbital motion were found to have time-variability and a dependence on the line transition from which the radial-velocity data was extracted. In particular, the deviations are strongly correlated with the top-level energy of the line transition. Physically this correlation may arise from the fact that lines with similar top-level energies originate from similar emission volumes in a system. The resulting hyperparameter vector for the covariance function is
\begin{equation}
\bm{\phi} = (A, l_1, l_2, \sigma_b)^T.
	\label{eq:covaraince_function_hyperparameter_vector}
\end{equation}

In Figure \ref{fig:gp_prior_condtional_and_mean} we display a visual representation of the GP framework described above. The figure is split into three columns with the full two-dimensional representation on the top row, and two one-dimensional slices in energy (15 eV and 25 eV) on the bottom row. The left-hand column shows a sample taken from the GP prior. The middle column shows a sample taken from the GP conditioned on a small amount of input data with no uncertainty. The right-hand column shows the same GP as the middle column but with the mean model -- the orbital velocities -- added back in. These plots were made with the hyperparameters fixed at $A=11^2 \kmpers$, $l_1=5^2$ days, $l_2=8^2$ eV, and $\sigma_w=0 \kmpers$ days. The plot shows how the squared exponential kernel specifies a prior over smoothly-varying functions, which after being given data, can learn the systematics present in a given system. The hyperparameters specify the nature of these systematics. For example, larger values of $A$ allow the GP to learn larger deviations in velocity from the mean model, and smaller values of $l_1$ or $l_2$ result in the GP learning more rapidly changing deviations in time and energy space respectively. Note that this is also how we generated benchmark dataset 3, previously described in Section \ref{sec:benchmark_dataset}, through randomly drawing different hyperparameter vectors to add a variety of different systematic problems into the orbital models. For the benchmark datasets the second dimension of the GP is actually redundant as, for simplicity, we focus on SB1 binaries with radial-velocity data from either one line only, or lines combined together into one input. However, in Section \ref{sec:discussion} we utilise the energy dimension and explore correlations between separate radial-velocity lines.

\begin{figure*}
	\includegraphics[width=\textwidth]{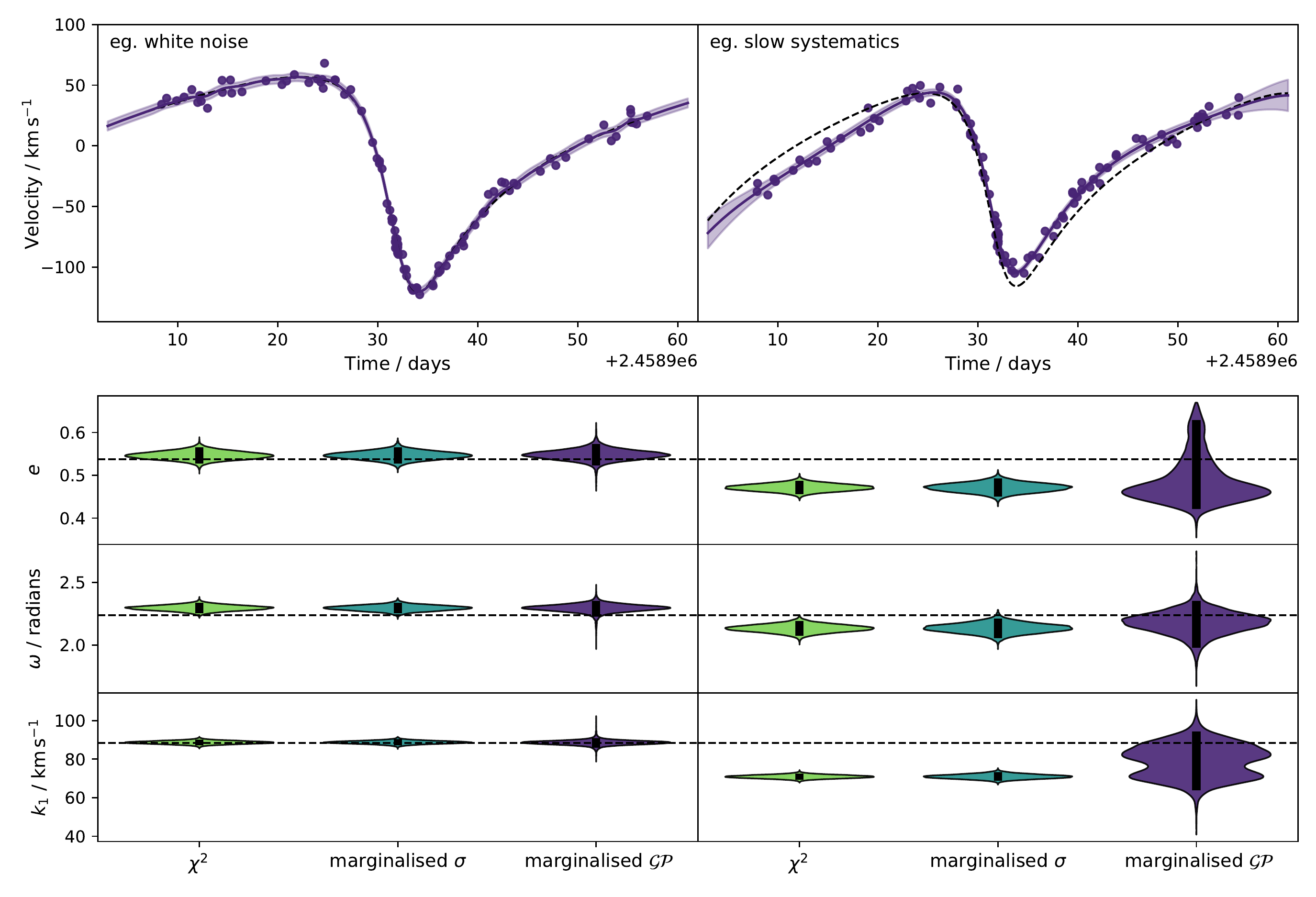}
    \caption{SB1 benchmark dataset example object 32/243. The left-hand column presents the object's radial-velocity data and model fits from benchmark dataset 1, containing white noise only. The right-hand column presents the corresponding object from dataset 3, now containing systematic noise on a similar timescale as the orbital period. In the top panels the dashed lines represent the mean models, while the solid purple represents the GP models and their $2\sigma$ uncertainties. In the bottom panels the dashed lines indicate the true values for each parameter, alongside their posterior parameter distributions for each of the three benchmark algorithms.}
    \label{fig:synthetic_sb1_example_violin_1}
\end{figure*}

\begin{figure*}
	\includegraphics[width=\textwidth]{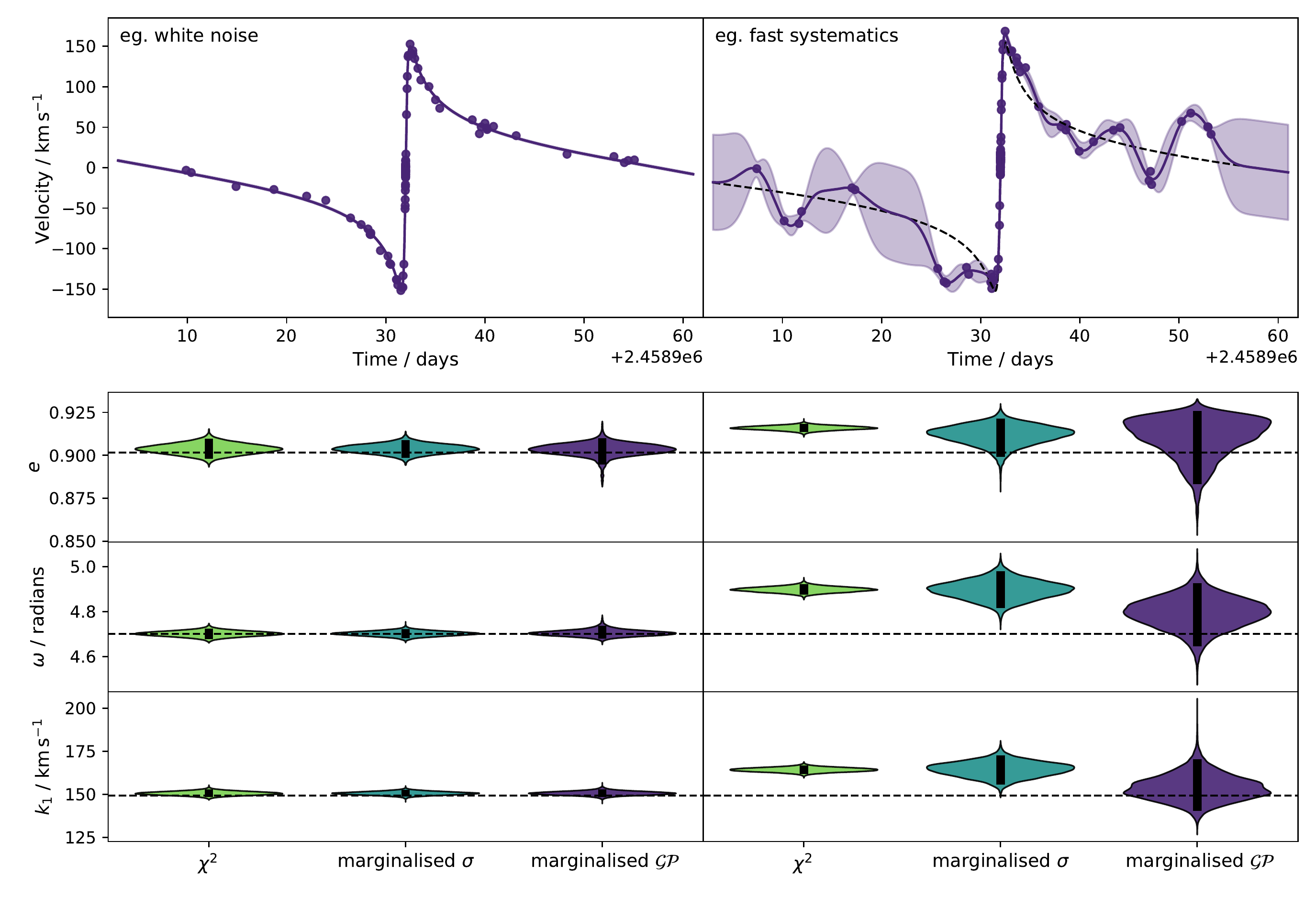}
    \caption{SB1 benchmark dataset example object 86/243. The same as Figure \ref{fig:synthetic_sb1_example_violin_1} except the object's systematics show variability on a much faster timescale.}
    \label{fig:synthetic_sb1_example_violin_2}
\end{figure*}

\subsection{Bayesian regression}
\label{subsec:bayesian_regression}
In order to perform Bayesian parameter estimation of the orbital parameters of interest, $\bm{\theta}$, we first define an objective function. The log-likelihood of the GP model is
\begin{equation}
\ln p(\bm{y} \mid \bm{X}, \bm{\theta}, \bm{\phi}) = -\frac{1}{2}\bm{r}\bm{K}^{-1}\bm{r} -\frac{1}{2} \ln \begin{vmatrix} \bm{K} \end{vmatrix} - \frac{m}{2} \ln(2\pi),
	\label{eq:log_likelihood_gp}
\end{equation}
where $\bm{r}=\bm{y} - \bm{v}$ is the residual vector after the data has had the mean model subtracted, and $m$ is the number of measurements. In this expression the first term accounts for the goodness-of-fit: as the residuals become smaller the value increases for a given covariance matrix. The second term acts as a complexity penalty. For perfect correlation (ie. no GP complexity) the determinant equals zero. As the correlation decreases, by shortening the characteristic length scales and/or increasing the amplitude, the determinant increases, the whole term decreases in value, and the extra complexity of the GP becomes less likely. The additional complexity can, however, be justified by a corresponding increase in the goodness-of-fit. Consequently, the GP will only have an impact on the model when the likelihood of the data having been generated solely by the mean orbital model is sufficiently low.

Following \citet[][]{Simpson2020MarginalisedSampling.}, we marginalise over both the hyperparameters of the covariance function as well as the free parameters in the mean model in order to best capture the predictive uncertainty in the parameters. We choose uniform priors over $\bm{\theta}$ and log-uniform priors over $\bm{\phi}$. We use the Gaussian process regression code, \texttt{george} \citep[][]{Ambikasaran2015FastProcesses}, and the affine-invariant Markov Chain Monte Carlo (MCMC) algorithm, \texttt{emcee}, implemented by \citet[][]{Foreman-Mackey2013Emcee:Hammer} to sample the probability distributions. We estimate the integrated autocorrelation time to ensure we draw sufficient samples for good convergence and to reduce the resulting errors in the posteriors down to the couple-of-percent level. For the other two benchmark algorithms, $\chi^2$ and marginalised $\sigma$, the regression is implemented in the exact same way except the covariance matrix is always diagonal and in the $\chi^2$ algorithm the white noise is fixed.

\section{Setting the standard}
\label{sec:setting_the_standard}
In Section \ref{sec:benchmark_dataset} we introduced three SB1 benchmark datasets, each with a different noise formulation. In Section \ref{sec:benchmark_algorithm} we described two commonly-employed algorithms, $\chi^2$ and marginalised $\sigma$, as well as proposing a novel application of a GP based algorithm, marginalised $\mathcal{GP}$. In this section we apply all three algorithms to all three benchmark datasets and report the results, setting the initial best scores to beat. The scoring is based on the criterion described in Section \ref{subsec:evaluation_criterion}, evaluating the predictive performance of the posterior parameter distribution when compared with the injected true values.

\begin{table}
  \centering
  \begin{tabular}{l c c c c c c}
  \hline
  Algorithm & \multicolumn{6}{c}{Dataset 1: white noise} \\
  & $T_0$ & $P$ & $e$ & $\omega$ & $k_1$ & $g_1$  \\
  \hline
  $\chi^2$ & \textbf{0.96} & \textbf{-1.30} & \textbf{3.55} & \textbf{2.30} & \textbf{-1.03} & \textbf{-1.35} \\
  marginalised $\sigma$ & 0.85 & -1.37 & 3.53 & 2.28 & -1.05 & -1.36\\
  marginalised $\mathcal{GP}$ & 0.83 & -1.41 & 3.51 & 2.24 & -1.12 & -1.45 \\
  \hline
  \end{tabular}
  \caption{Scores for dataset 1 for each of the three benchmark algorithms. Bold indicates the best scoring algorithm per parameter. The score is defined in Section \ref{subsec:evaluation_criterion} as the median logarithmic density of the predicted posterior distribution at the true value for each of the orbital parameters.}
    \label{tab:synethtic_benchmarks_white_noise}
\end{table}
\begin{table}
  \centering
  \begin{tabular}{l c c c c c c}
  \hline
  Algorithm & \multicolumn{6}{c}{Dataset 2: polynomial systematics} \\
  & $T_0$ & $P$ & $e$ & $\omega$ & $k_1$ & $g_1$  \\
  \hline
  $\chi^2$ & -1.84 & -19.1 & -0.77 & -3.34 & -6.62 & -25.1 \\
  marginalised $\sigma$ & -1.49 & -6.94 & 0.59 & -1.32 & -3.97 & -6.64\\
  marginalised $\mathcal{GP}$ & \textbf{0.13} & \textbf{-3.42} & \textbf{2.09} & \textbf{0.89} & \textbf{-3.00} & \textbf{-3.83} \\
  \hline
  \end{tabular}
  \caption{The same as Table \ref{tab:synethtic_benchmarks_white_noise} but for dataset 2.}
    \label{tab:synethtic_benchmarks_ploynomial_systemtics}
\end{table}
\begin{table}
  \centering
  \begin{tabular}{l c c c c c c}
  \hline
  Algorithm & \multicolumn{6}{c}{Dataset 3: $\mathcal{GP}$ systematics} \\
  & $T_0$ & $P$ & $e$ & $\omega$ & $k_1$ & $g_1$  \\
  \hline
  $\chi^2$ & -3.99 & -76.7 & -0.45 & -5.81 & -486 & -3130 \\
  marginalised $\sigma$ & -1.89 & -6.63 & 1.16 & -1.92 & -8.48 & -24.9\\
  marginalised $\mathcal{GP}$ & \textbf{0.01} & \textbf{-3.71} & \textbf{2.04} & \textbf{0.96} & \textbf{-3.57} & \textbf{-4.49} \\
  \hline
  \end{tabular}
  \caption{The same as Table \ref{tab:synethtic_benchmarks_white_noise} but for dataset 3.}
    \label{tab:synethtic_benchmarks_gp_systematics}
\end{table}

In Tables \ref{tab:synethtic_benchmarks_white_noise}, \ref{tab:synethtic_benchmarks_ploynomial_systemtics}, and \ref{tab:synethtic_benchmarks_gp_systematics} we present the results for benchmark datasets 1, 2, and 3 respectively. We find that our marginalised $\mathcal{GP}$ algorithm substantially outperforms the other more traditional algorithms in both datasets 2 and 3, ie. for those with non-zero systematics. The performance is comparable between these two datasets' scores indicating that they are set at a similar difficulty level. For dataset 1, all three algorithms perform very similarly, although the $\chi^2$ algorithm narrowly gives the best scores. These scores demonstrate how both the marginalised $\sigma$ and marginalised $\mathcal{GP}$ algorithms (almost) reduce to the $\chi^2$ algorithm in the case of purely white noise.

In Figures \ref{fig:synthetic_sb1_example_violin_1} and \ref{fig:synthetic_sb1_example_violin_2} we display two illustrative examples from the benchmark results. In both Figures the left-hand column shows a system from dataset 1, containing white noise only, whilst the right-hand column shows the corresponding system from dataset 3, now contaminated by systematic noise. In Figure \ref{fig:synthetic_sb1_example_violin_1} the systematics are varying on roughly the same timescale as the orbital period and in Figure \ref{fig:synthetic_sb1_example_violin_2} they vary much faster with multiple mean model crossings per orbital period. In the top panels we plot the marginalised $\mathcal{GP}$ models along with their $2\sigma$ uncertainties. In the bottom panels we show violin plots of the posterior distributions for three of the orbital parameters, $e$, $\omega$, and $k_1$, relative to the true values (dashed lines). In the regime of purely white noise we see how all three models accurately and confidently predict the true parameter values. The marginalised $\mathcal{GP}$ algorithm does have slightly longer tails in its predictive distributions, which manifests itself as an ever so slightly lower score in Table \ref{tab:synethtic_benchmarks_white_noise}. In the regime of systematic noise we can immediately see how the marginalised $\mathcal{GP}$ algorithm predicts more complex and uncertain posterior distributions in each of the orbital parameters. Consequently, this algorithm is far more accurate in recovering the true value, whilst the other two algorithms are overconfident given their inaccuracy.

The GP algorithm outperforms the other algorithms when systematic noise is present in the data. The $\chi^2$ and marginalised $\sigma$ algorithms return the orbit that minimises their velocity residuals, which is not necessarily representative of the true underlying orbit if the systematics are asymmetrically distributed about the mean model. Conversely, the GP algorithm can generate flexible systematic functions, so long as the systematics can be disentangled from the mean model. In general this is the case, as is shown in Figures \ref{fig:synthetic_sb1_example_violin_1} and \ref{fig:synthetic_sb1_example_violin_2}, although the GP model will still of course struggle if the mean model plus the systematics result in radial velocities that can be confused with a different mean model parameterisation. But for the most part, the combinations of various orbital models and GP systematics are marginalised over and the resulting parameter posterior distributions incorporate this uncertainty, making the true values more likely to be recovered by this algorithm.

\section{Discussion}
\label{sec:discussion}
We have presented benchmark datasets and algorithms for the orbits of emission-line binaries. We have found that when systematics are present in the data, simply inflating the uncertainties does not always modify the predicted orbital parameter distributions enough to recover the true values. Therefore, we find utilising our marginalised $\mathcal{GP}$ algorithm is important for predicting accurate orbital parameter distributions. This has important implications for the derived orbits of emission-line stars, their dynamical masses, and the binary evolutionary scenarios that may be inferred from radial-velocity data.

The generative nature of the GP algorithm has further advantages, such that we can separate out orbital motion from other potential physical signals lurking in the data. The GP model can then be used to help identify the sources of systematics and further characterise different aspects of the system, rather than sweeping these signals under the carpet by assuming the uncertainties in the data are underestimated.

\subsection{The performance impact of additional data}
\label{subsec:performance_impact_of_additional_data}
In this section we analyse how having additional data -- such as radial velocities from different spectral lines or stellar components, or having relative astrometric measurements -- can impact the predictive distributions of our marginalised $\mathcal{GP}$ algorithm. To facilitate the discussion we set a baseline score for all the tests using an SB1 object with orbital parameters $P=50$ days, $T_0=2458932$ (JD), $e=0.3$, $\omega=270^\circ$, $k_1=50 \kmpers$, and $g_1=0 \kmpers$. The primary star's radial velocities include white noise at the $5 \kmpers$ level and systematics drawn from a GP with a squared exponential kernel, parameterised by $A_1=20^2 \kmpers$ and $l_1=2^2$ days.

\subsubsection{Multiple spectroscopic lines}
\label{subsubsec:multiple_spectroscopic_lines}
Often radial velocities are extracted from many spectral lines and combined, or extracted from entire portions of the spectra, resulting in one radial-velocity curve for a given binary system. However, it may be the case that different spectral lines display different systematic behaviour; and therefore, it is preferable to treat the systematics from each line individually. To test the effect of utilising multiple spectral lines on our marginalised $\mathcal{GP}$ algorithm we create a test grid which varies the number of spectral lines and the amplitude of systematics present. Each spectral line has identical underlying orbital parameters, as well as systematics drawn from the prior distribution of a GP parameterised as per the baseline object, only with different values of $A_1$. This dataset spans incremental increases in the number of spectral lines from 1 to 5 and systematic amplitudes linearly increasing over the interval $0 \kmpers < A_1 < 20^2 \kmpers$, resulting in a 5x5 grid.

\begin{figure*}
	\includegraphics[width=\textwidth]{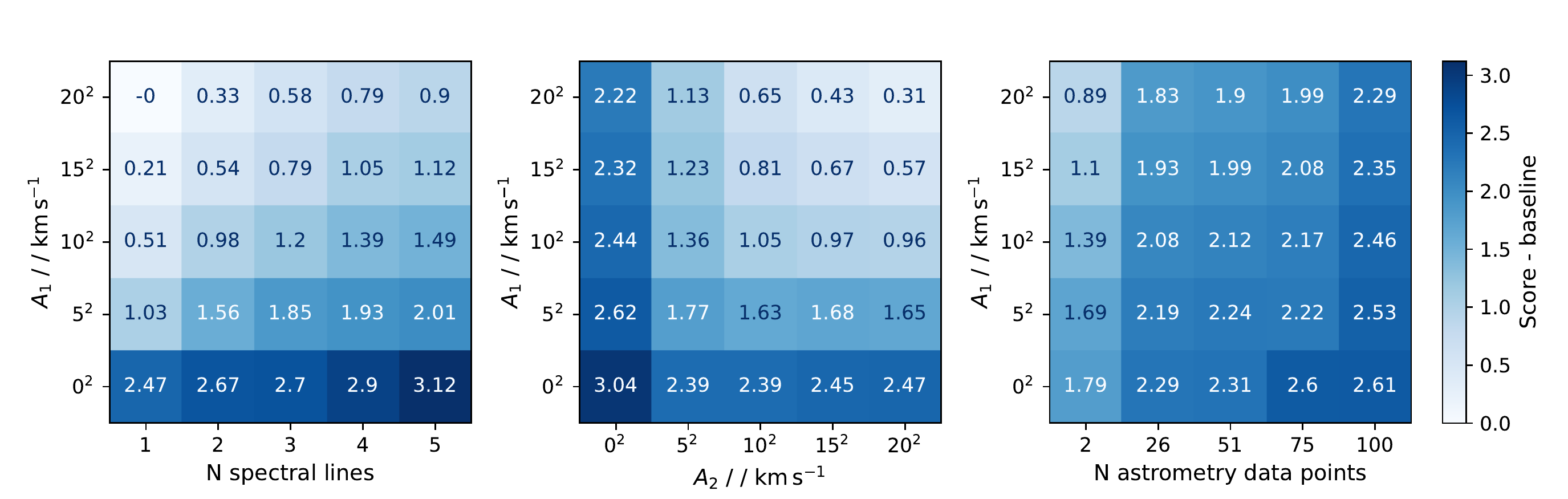}
    \caption{ The impact of additional data on the performance of our marginalised $\mathcal{GP}$ algorithm. The left-hand panel tests the effect of using multiple spectroscopic lines, the middle panel tests the effect of having SB2 data of varying quality, and the right-hand panel tests the effect of having complementary relative astrometry data. In all three tests we also vary the level of systematics in the primary radial velocities, $A_1$. The grid colours and values represent the score, as defined in Section \ref{subsec:evaluation_criterion}, relative to the score of the baseline object as defined in Section \ref{subsec:performance_impact_of_additional_data}. NB. the baseline object is the same as the  test run in the top-left grid point of the left-hand panel, hence a score of zero.}
    \label{fig:additional_data_impact_discussion}
\end{figure*}

We run the marginalised $\mathcal{GP}$ algorithm over the test grid, and in this case we utilise the second dimension of the GP, allowing the model to learn systematic correlations in the energy dimension. The results are presented in the left-hand panel of Figure \ref{fig:additional_data_impact_discussion}. Here we show the scores, as defined in Section \ref{subsec:evaluation_criterion}, relative to the baseline object. These scores have been averaged across the 6 orbital parameters. As such, positive values reflect an improvement in the inference of the orbital parameters relative to the baseline, owing to the inclusion of the additional data. Note that the baseline object has the same parameterisation as the test run in the top-left grid point, hence a score of zero. We find that the more spectral lines included in the parameter estimation, the better the score becomes. Once all 5 lines are utilised, the scores increase by values between 0.65 and 0.98 depending on the level of systematics present.

\subsubsection{Double-lined spectroscopic binaries}
\label{subsubsec:double_lined_Spectroscopic_binaries}
Spectra of emission-line binaries may include lines from both stellar components, known as a double-lined spectroscopic binary (SB2). In this case it is possible to co-fit both radial-velocity curves with the same orbital model as most of the parameters are shared. To test the effect of having SB2 data on our marginalised $\mathcal{GP}$ algorithm we create a test grid which varies the amplitude of systematics present in both the primary's and secondary's radial-velocity data. Both stellar components have identical underlying orbital parameters except for the secondary's semi-amplitude which is set at $k_2=100 \kmpers$. The systematics are drawn from the prior distribution of a GP parameterised as per the baseline object, only with different values of $A_1$ and $A_2$. This dataset spans systematic amplitudes linearly increasing over the interval $0 \kmpers < A_1 < 20^2 \kmpers$ and $0 \kmpers < A_2 < 20^2 \kmpers$, resulting in a 5x5 grid. For $A_2 = 0 \kmpers$ this simulates the scenario when the secondary is, for example, an O-star with easy to extract absorption lines, and therefore little to no systematics in the radial velocities. For $A_2 = 20^2 \kmpers$ this simulates having another emission-line star where substantial systematics may remain in the data.

We run the marginalised $\mathcal{GP}$ algorithm over the test grid. The results are presented in the middle panel of Figure \ref{fig:additional_data_impact_discussion}. Here we show the scores relative to the baseline object, averaged across the 6 orbital parameters. We find that for small or no systematics in the secondary's radial velocities the score is improved on average by 2.22 across the 6 orbital parameters. As the amplitude of the secondary's systematics are increased, the score decreases. Once the amplitude of the secondary's systematics are comparable to the primary's, the average score has returned to similar values to that of the SB1 case. Additionally, as the amplitude of the primary's systematics are decreased, the scores show a corresponding increase as expected. We also find that the lower-right triangle of the grid slightly outperforms the opposite half. This result is due to the secondary having a larger radial-velocity semi-amplitude than the primary, and therefore systematics of the same absolute amplitude appear as less of a nuisance to the secondary's data. Overall, we find that the presence of the secondary's data helps the algorithm to distinguish between the mean orbital model and the systematics, especially when the secondary itself is reasonably unaffected by systematics.

\subsubsection{Relative astrometry}
\label{subsubsec:relative_astrometry}
For some (close by) systems it is possible to observe the relative motion of the binary stars through precise relative astrometry from interferometric images. Through combining the radial velocities and relative astrometric data we can constrain the entire three-dimensional geometry of the orbits. To test the effect of having relative astrometry data in addition to SB1 data we create a test grid which varies the number of astrometric observations available and the amplitude of systematics present in the radial-velocity data. Each test has identical underlying orbital parameters and systematics as per the baseline object. This dataset spans incremental increases in the number of astrometric observations available from 2 to 100 and systematic amplitudes linearly increasing over the interval $0 \kmpers < A_1 < 20^2 \kmpers$, resulting in a 5x5 grid. The relative astrometry data has white noise added at the 5\% level, and the observations are sampled from a uniform distribution over the same time interval as the radial velocities.

We run the marginalised $\mathcal{GP}$ algorithm over the test grid. The results are presented in the right-hand panel of Figure \ref{fig:additional_data_impact_discussion}. Here we show the scores relative to the baseline object, averaged across the 6 orbital parameters. We find even having as little as two astrometric observations helps to substantially improve the average score of the orbital parameter inference. In fact, for challenging systematics in the radial velocities, such as when $A_1 = 20^2 \kmpers$, having 26 astrometry data points results in better parameter inference than a dataset with no systematics ($A_1 = 0^2 \kmpers$). The scores continue to improve further across the grid as the number of astrometric observations increases. The parameters $P$, $T_0$, $e$, and $\omega$ get the largest boost in performance as they appear directly in the astrometric orbital model, but interestingly the parameters $k_1$ and $g_1$ also receive a modest improvement, probably as an indirect consequence of the other parameter distributions becoming more accurate.

\subsection{Application to prototypical systems}
\label{subsec:application_to_prototypical}
In this section we apply our best benchmark algorithm, the marginalised $\mathcal{GP}$, to some real datasets of emission-line binaries to compare how our predicted parameter distributions differ from those previously published and to elucidate systematics that may otherwise go untraced.

\subsubsection{Eta Carinae}
\label{subsubsec:eta_carinae}

Eta Carinae is a long-period (2022.7 days \citep[][]{Damineli2008TheEvents}), highly eccentric ($e\sim0.9$) SB1 system situated at a distance of $2.3 \pm 0.1 \kpc$ \citep[][]{Allen1993TheCarinae, Smith2006TheCarinae}. The primary star is a luminous blue variable with an exceptionally strong wind; it has a mass-loss rate of $8.5 \times 10^{-4} \masslossrate$ and terminal wind velocity of $420 \kmpers$ \citep[][]{Hillier2001, Groh2012OnSpectra, Clementel20153DApastron}. The orbital parameters were originally calculated by \citet[][]{Damineli1997EtaBinary}. Most recently \citet[][]{Grant2020Uncovering140} re-computed the orbital parameters using the Balmer lines and the mean model described in equation \ref{eq:convolutional_keplerian_mean_model}.

\begin{table}
    \centering
    \renewcommand{\arraystretch}{1.3}
    \begin{tabular}{l c c}
    \hline
    Eta Carinae's orbit & \citet[][]{Grant2020Uncovering140} & marginalised $\mathcal{GP}$ \\
    \hline
    \multicolumn{3}{c}{Elements of the system} \\
    \hline
    $P$ (fixed) & $2022.7$ & $2022.7$ \\
    $T_0$ & $2454848.3^{+0.4}_{-0.4}$ & $2454850.1^{+2.3}_{-2.6}$ \\
    $e$ & $0.91^{+0.003}_{-0.003}$ & $0.90^{+0.01}_{-0.01}$ \\
    $\omega$ & $241^{+1}_{-1}$ & $248^{+4}_{-5}$ \\
    \hline
    \multicolumn{3}{c}{Elements of the radial-velocity orbit} \\
    \hline
    $k_1$ & $69.0^{+0.9}_{-0.9}$ & $66.8^{+4.6}_{-3.9}$ \\
    $\gamma_1$ (fixed) & - & - \\
    \hline
    \end{tabular}
    \caption{Results for Eta Carinae comparing literature parameter distributions, using the $\chi^2$ algorithm (middle), and our estimated parameter distributions, using our marginalised $\mathcal{GP}$ algorithm (right). The uncertainties are the $68$\% credible intervals. The orbital parameters are the period, $P$ (days), the time of periastron, $T_0$ (JD), the eccentricity, $e$, the argument of periastron, $\omega$ (degrees), the primary star's semi-amplitude, $k_1$ ($\rm{\,km \, s^{-1}}$), and the primary star's radial-velocity offset, $\gamma_1$ ($\rm{\,km \, s^{-1}}$).}
    \label{tab:application_to_prototypical_systems_eta_car}
\end{table}

\begin{figure}
	\includegraphics[width=1\columnwidth]{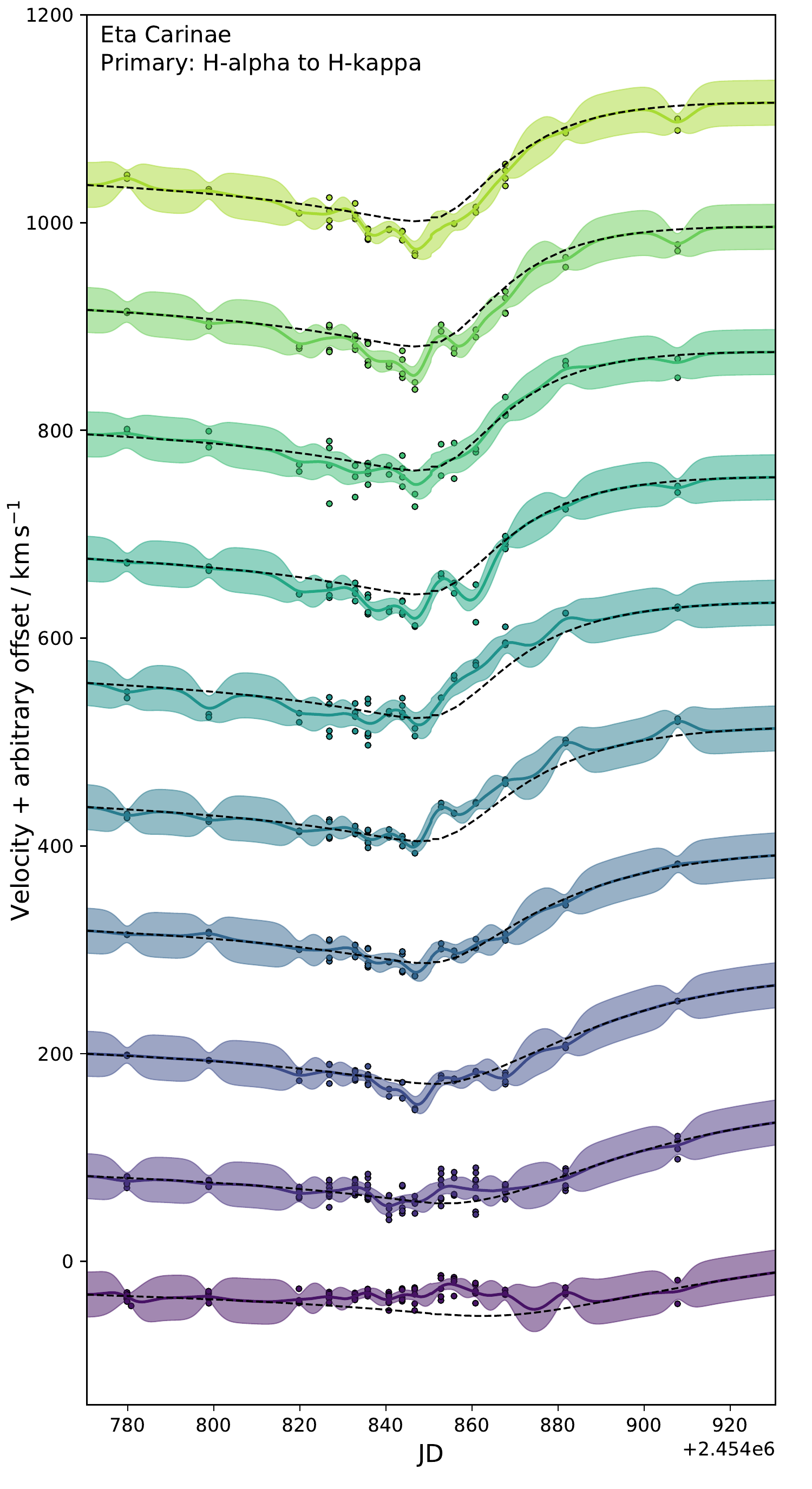}
    \caption{Marginalised $\mathcal{GP}$ algorithm conditioned on Eta Carinae's observed data. The mean model is shown as the black dashed line, the GP is shown at sequential spectral line energies by the coloured lines from H-alpha (purple) to H-kappa (yellow), and each shaded region represents the $2\sigma$ uncertainty in the GP. The $\gamma$ values are arbitrary.}
    \label{fig:eta_car_gp}
\end{figure}

We utilise the same spectral data as \citet[][]{Grant2020Uncovering140} which is from an open-source online archive for Eta Carinae\footnote{\url{http://etacar.umn.edu/archive/}}. The dataset contains between $57$ and $132$ radial-velocity measurements per Balmer line. The Balmer lines included span H-alpha to H-kappa with corresponding top-level line-transition energies from $12.09$ eV to $13.50$ eV.

We first apply the $\chi^2$ algorithm to the data and check that we recover predicted parameter distributions that are consistent with \citet[][]{Grant2020Uncovering140}. We note that similarly to \citet[][]{Grant2020Uncovering140} we fix the orbital period and pre-set the $\gamma$ values for each spectral line. Next, we apply our marginalised $\mathcal{GP}$ algorithm to the same data and report the results in Table \ref{tab:application_to_prototypical_systems_eta_car}. We find that the distributions of all the parameters become significantly more uncertain when using our marginalised $\mathcal{GP}$ algorithm.

\begin{figure}
	\includegraphics[width=1\columnwidth]{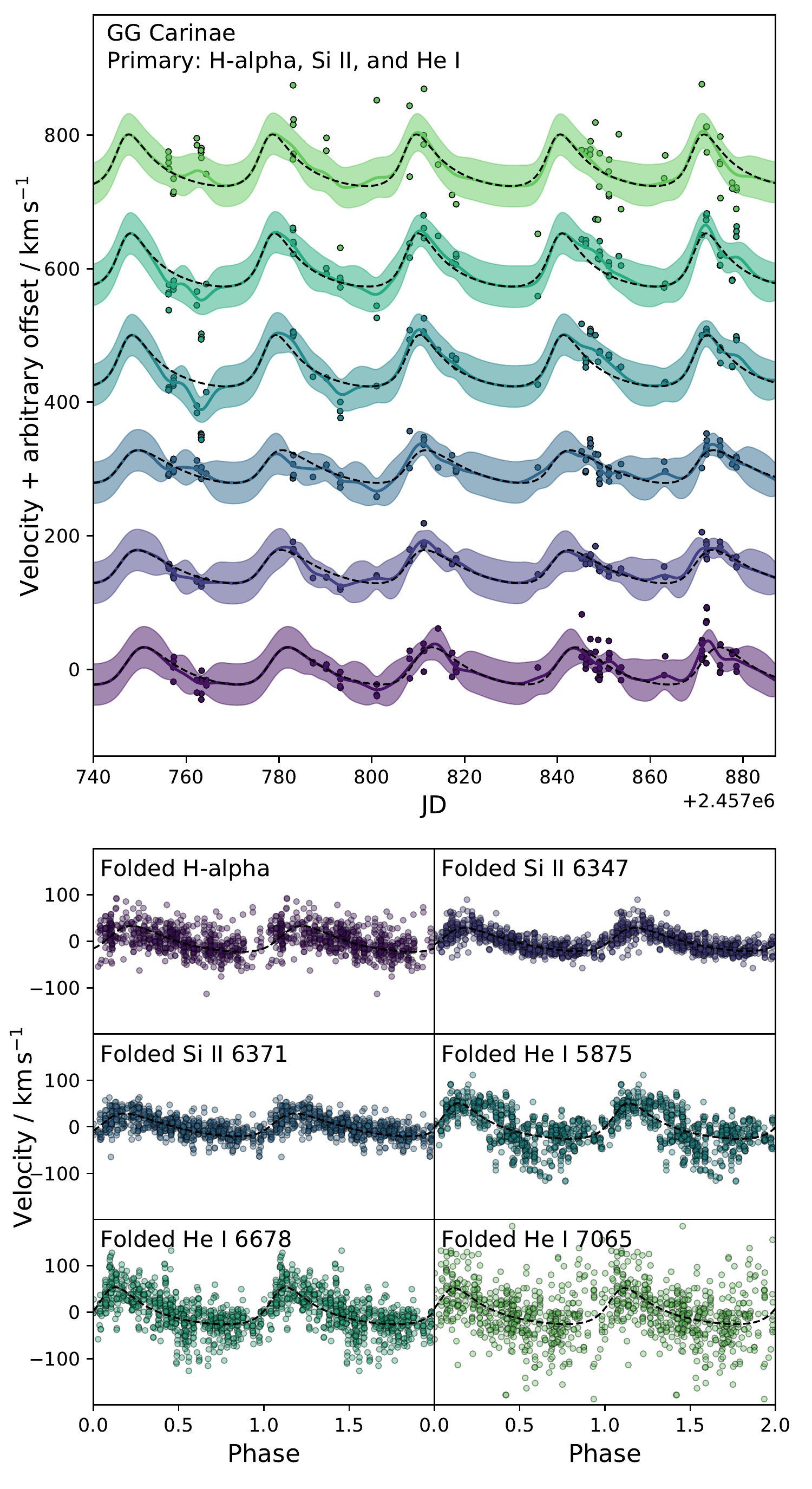}
    \caption{Top panel: a 147-day snapshot of the marginalised $\mathcal{GP}$ algorithm conditioned on GG Carinae's observed data. The mean model is shown as the black dashed line, the GP is shown at sequential spectral line energies by the coloured lines from H-alpha (purple) to He I (green), and each shaded region represents the $2\sigma$ uncertainty in the GP. The $\gamma$ values are arbitrary. Bottom panel: all the radial-velocity data folded by the predicted period.}
    \label{fig:gg_car_gp}
\end{figure}

In Figure \ref{fig:eta_car_gp} we display the conditioned GP for Eta Carinae. In this fit we can see the data exhibit obvious systematics, the most prominent of which occur in H-alpha. Despite this the GP is able to fully capture the variability of the data about the mean model. The median characteristic length scale of the marginalised GP kernel is $10.4$ days which manifests itself as systematics varying on timescales much faster than the orbital period.

The overall shift to lower values of $k_1$, for our predicted distribution relative to the literature values, results in the predicted mass ratio, $q=m_2/m_1$, moving to lower values. Therefore the extent to which the primary star's mass is greater than the secondary's may be larger than previously inferred.

\subsubsection{GG Carinae}
\label{subsubsec:gg_carinae}

\begin{table}
    \centering
    \renewcommand{\arraystretch}{1.3}
    \begin{tabular}{l c c}
    \hline
    GG Carinae's orbit & \citet[][]{Porter2021GGPhotometry} & marginalised $\mathcal{GP}$ \\
    \hline
    \multicolumn{3}{c}{Elements of the system} \\
    \hline
    $P$ & $31.01^{+0.01}_{-0.01}$ & $30.99^{+0.02}_{-0.02}$ \\
    $T_0$ & $2452069.36^{+1.30}_{-1.30}$ & $2452071.98^{+2.94}_{-2.93}$ \\
    $e$ & $0.50^{+0.03}_{-0.03}$ & $0.50^{+0.06}_{-0.06}$ \\
    $\omega$ & $339.87^{+3.10}_{-3.06}$ & $334.36^{+6.10}_{-6.10}$ \\
    \hline
    \multicolumn{3}{c}{Elements of the radial-velocity orbit} \\
    \hline
    $k_1$ & $48.57^{+2.04}_{-1.87}$ & $49.48^{+3.93}_{-3.44}$ \\
    $\gamma_1$ & $-0.72^{+0.36}_{-0.36}$ & $-0.41^{+0.47}_{-0.48}$ \\
    \hline
    \end{tabular}
    \caption{The same as Table \ref{tab:application_to_prototypical_systems_eta_car} except for GG Carinae.}
    \label{tab:application_to_prototypical_systems_gg_car}
\end{table}

GG Carinae is an SB1 system with a ${\sim}31$ day period \citep[][]{Hernandez1981FurtherCarinae, Gosset1985RadialCarinae, Marchiano2012TheCarinae} at a distance of $3.4^{+0.7}_{-0.5} \kpc$ \citep[][]{GaiaCollaboration2016TheMission, GaiaCollaboration2018GaiaProperties}. The primary star is a B[e] supergiant with a mass-loss rate of $2.2 \times 10^{-6} \masslossrate$ and terminal wind velocity of $265 \kmpers$ \citep[][]{Porter2021GGPhotometry}. We use the orbital parameters calculated by \citet[][]{Porter2021GGPhotometry} using the same modelling techniques as used for Eta Carinae, making use of the mean model described in equation \ref{eq:convolutional_keplerian_mean_model}.

We utilise the same spectral data as \citet[][]{Porter2021GGPhotometry} which is from the Global Jet Watch network of telescopes. The dataset contains between $513$ and $718$ radial-velocity measurements per spectral line. The lines included are H-alpha, Si II $6347$ \AA, Si II $6371$ \AA, He I $5875$ \AA, He I $6678$ \AA, and He I $7065$ \AA{} with corresponding top-level line-transition energies from $12.09$ eV to $23.07$ eV.

We first apply the $\chi^2$ algorithm to the data and check that we recover predicted parameter distributions that are consistent with \citet[][]{Porter2021GGPhotometry}. Similarly to \citet[][]{Porter2021GGPhotometry} we pre-set the $\gamma$ values for each spectral line, but still fit for an overall offset. Next, we apply our marginalised $\mathcal{GP}$ algorithm to the same data and report the results in Table \ref{tab:application_to_prototypical_systems_gg_car}. We find that the distributions of all the parameters become more uncertain, by approximately a factor of two, when using our marginalised $\mathcal{GP}$ algorithm.

\begin{table}
    \centering
    \renewcommand{\arraystretch}{1.3}
    \begin{tabular}{l c c}
    \hline
    WR 140's orbit & \citet[][]{Thomas2021The140} & marginalised $\mathcal{GP}$ \\
    \hline
    \multicolumn{3}{c}{Elements of the system} \\
    \hline
    $P$ & $2895.00^{+0.29}_{-0.29}$ & $2896.34^{+0.67}_{-0.64}$ \\
    $T_0$ & $2460636.73^{+0.53}_{-0.53}$ & $2460640.45^{+1.29}_{-1.33}$ \\
    $e$ & $0.8993^{+0.0013}_{-0.0013}$ & $0.8980^{+0.0029}_{-0.0030}$ \\
    $\omega_{\scalebox{0.6}{\rm{WR}}}$ & $227.44^{+0.52}_{-0.52}$ & $228.20^{+1.16}_{-1.18}$ \\
    \hline
    \multicolumn{3}{c}{Elements of the radial-velocity orbit} \\
    \hline
    $k_{\scalebox{0.6}{\rm{WR}}}$ & $75.25^{+0.63}_{-0.63}$ & $72.70^{+2.02}_{-2.04}$ \\
    $k_{\scalebox{0.6}{\rm{O}}}$ & $26.50^{+0.48}_{-0.48}$ & $26.32^{+1.25}_{-1.29}$ \\
    $\gamma_{\scalebox{0.6}{\rm{WR}}}$ & $0.26^{+0.32}_{-0.32}$ & $0.98^{+1.16}_{-1.14}$ \\
    $\gamma_{\scalebox{0.6}{\rm{O}}}$ & $3.99^{+0.37}_{-0.37}$ & $2.32^{+1.00}_{-1.15}$ \\
    \hline
    \multicolumn{3}{c}{Elements of the astrometric orbit} \\
    \hline
    $a$ & $8.922^{+0.067}_{-0.067}$ & $9.011^{+0.117}_{-0.112}$ \\
    $i$ & $119.07^{+0.88}_{-0.88}$ & $119.09^{+0.89}_{-0.87}$ \\
    $\Omega$ & $353.87^{+0.67}_{-0.67}$ & $354.43^{+0.88}_{-0.87}$ \\
    \hline
    \end{tabular}
    \caption{Results for WR 140 comparing literature parameter distributions, using the $\chi^2$ algorithm (middle), and our estimated parameter distributions, using our marginalised $\mathcal{GP}$ algorithm (right). The uncertainties are the $68$\% credible intervals. The orbital parameters are the period, $P$ (days), the time of periastron, $T_0$ (JD), the eccentricity, $e$, the argument of periastron of the Wolf-Rayet star, $\omega_{\scalebox{0.6}{\rm{WR}}}$ (degrees), the Wolf-Rayet's and O-star's semi-amplitudes, $k_{\scalebox{0.6}{\rm{WR}}}$ and $k_{\scalebox{0.6}{\rm{O}}}$ ($\rm{\,km \, s^{-1}}$), the Wolf-Rayet's and O-star's radial-velocity offsets, $\gamma_{\scalebox{0.6}{\rm{WR}}}$ and $\gamma_{\scalebox{0.6}{\rm{O}}}$ ($\rm{\,km \, s^{-1}}$), the semi-major axis, $a$ (mas), the inclination, $i$ (degrees), and the longitude of the ascending node, $\Omega$ (degrees).}
    \label{tab:application_to_prototypical_systems_wr_140}
\end{table}

In the top panel of Figure \ref{fig:gg_car_gp} we display a snapshot of the conditioned GP for GG Carinae, and in the bottom panel we show all of the radial-velocity data folded by the derived period. In this fit we can see the data exhibits sporadic systematic offsets from the orbital motion. The GP is able to capture the systematics, and in doing so, the predictions for the orbital parameters become more uncertain. The median characteristic length scale of the marginalised GP kernel is $3.9$ days which manifests itself as systematics varying on timescales faster than the orbital period.

The distributions for the literature and our prediction for $k_1$ are broadly consistent, however the overall shift to higher values of $k_1$, for our predicted distribution, results in the predicted mass ratio, $q=m_2/m_1$, moving to higher values. Therefore the extent to which the primary star's mass is greater than the secondary's may be smaller than previously predicted.

\subsubsection{WR 140}
\label{subsubsec:wr_140}

\begin{figure}
	\includegraphics[width=1\columnwidth]{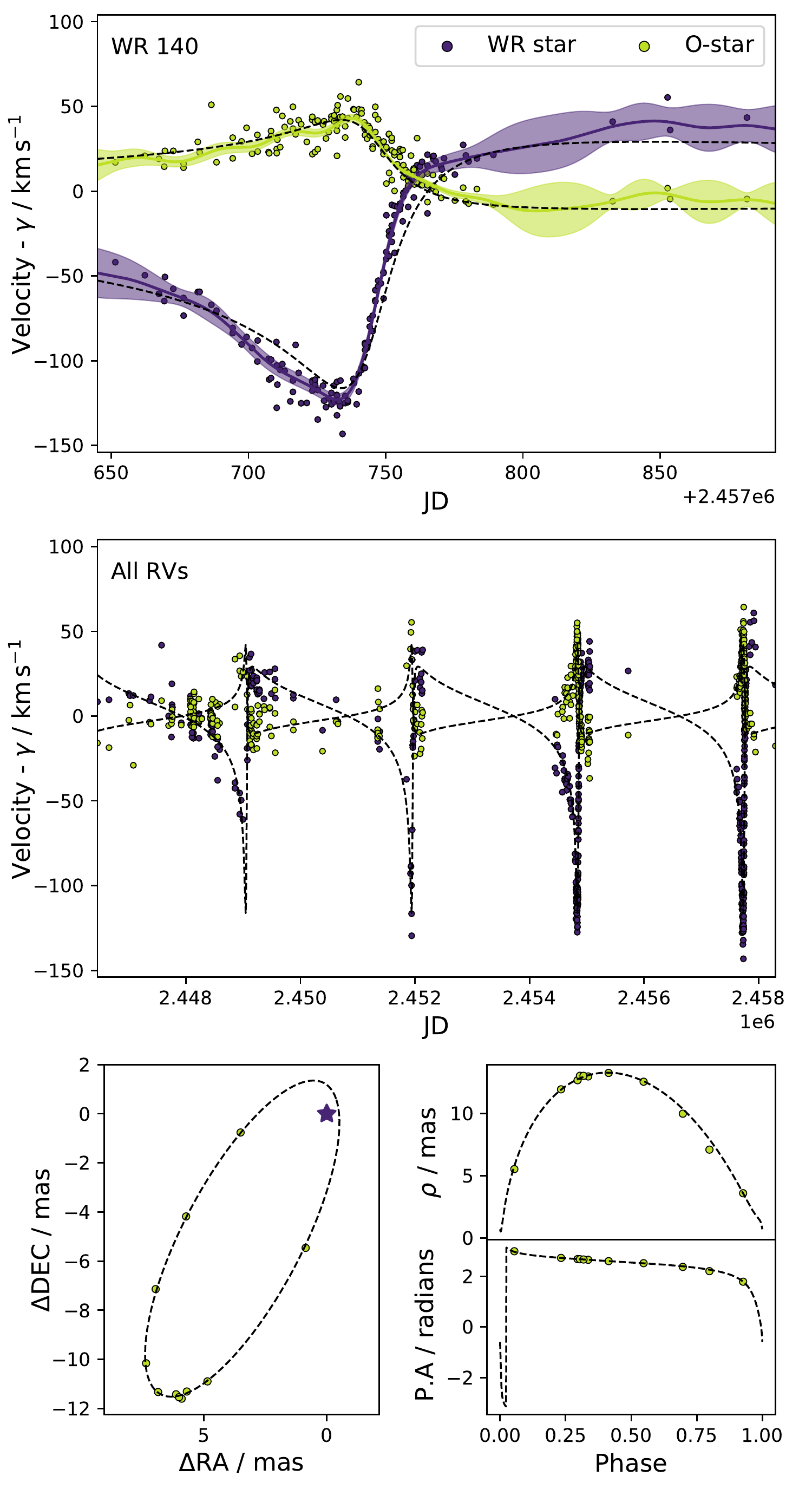}
    \caption{Top panel: a 247-day snapshot of the marginalised $\mathcal{GP}$ algorithm conditioned on WR 140's observed data. The mean models are shown as the black dashed lines, the GPs are shown for the Wolf-Rayet star (purple) and the O-star (yellow), and each shaded region represents the $2\sigma$ uncertainty in the GPs. The $\gamma$ values have been subtracted. Middle panel: all the radial-velocity data used in the parameter estimation. Bottom panels: the relative astrometric data and the best-fit model, with the Wolf-Rayet star held fixed at the origin. $\rho$ is the stellar separation and P.A. is the position angle.}
    \label{fig:wr_140_gp}
\end{figure}

\begin{figure}
	\includegraphics[width=1\columnwidth]{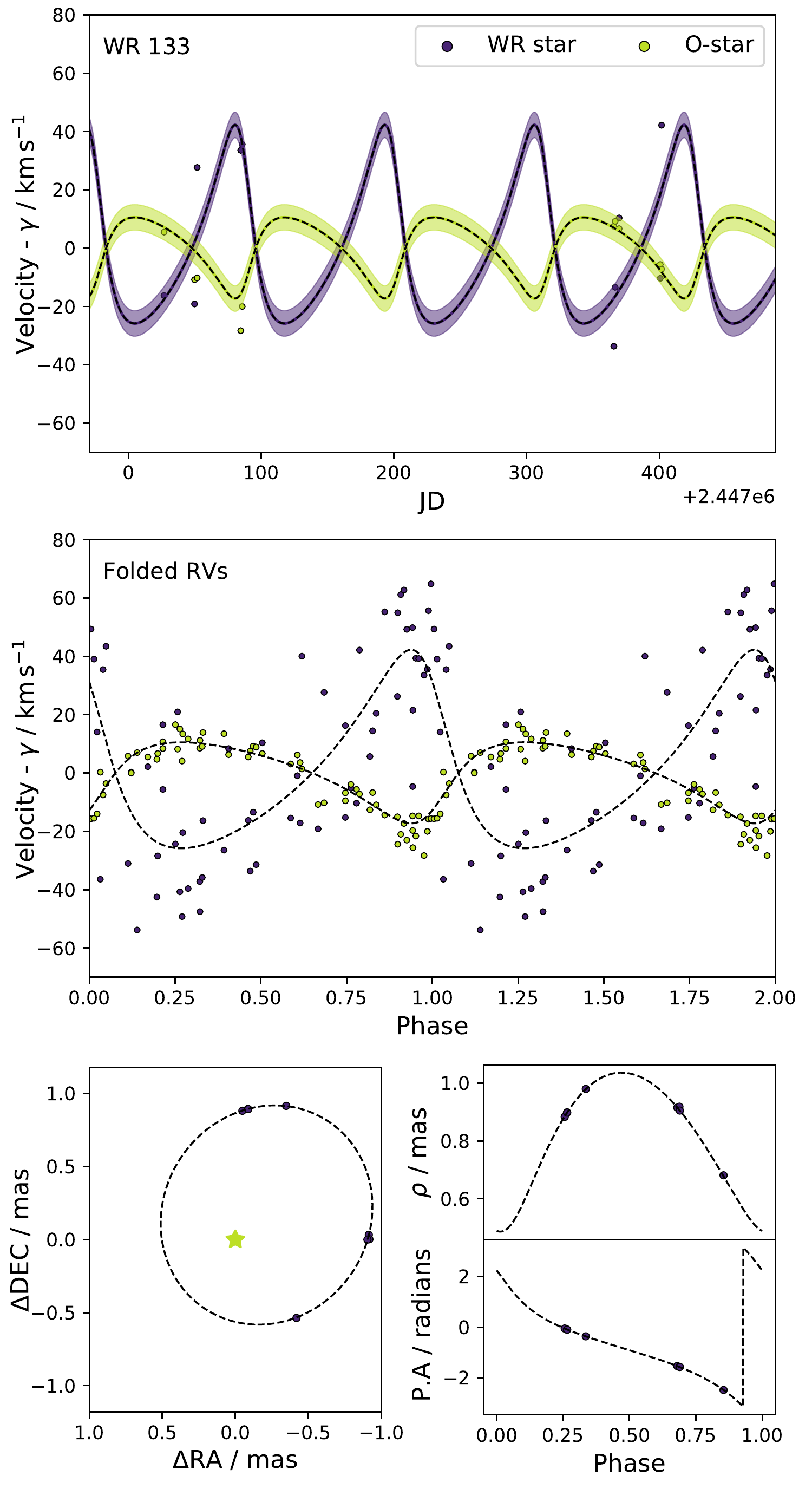}
    \caption{Top panel: a 517-day snapshot of the marginalised $\mathcal{GP}$ algorithm conditioned on WR 133's observed data. The mean models are shown as the black dashed lines, the GPs are shown for the Wolf-Rayet star (purple) and the O-star (yellow), and each shaded region represents the $2\sigma$ uncertainty in the GPs. The $\gamma$ values have been subtracted. Middle panel: all the radial-velocity data folded by the predicted period. Bottom panels: the relative astrometric data and the best-fit model, with the O-star held fixed at the origin. $\rho$ is the stellar separation and P.A. is the position angle.}
    \label{fig:wr_133_gp}
\end{figure}

WR 140 is a long-period (7.992 years), highly eccentric ($0.9$) SB2 and visual system situated at a distance of $1.64^{+0.08}_{-0.07} \kpc$ \citep[][]{GaiaCollaboration2016TheMission, GaiaCollaboration2018GaiaProperties, Bailer-Jones2018Estimating2}. The stellar components are a carbon-rich Wolf-Rayet star (WC7pd) and an O-star (O5.5fc). We use the orbital parameters recently calculated by \citet[][]{Thomas2021The140} as the current literature estimates.

We utilise the same data as \citet[][]{Thomas2021The140}: radial-velocity measurements from \citet[][]{Marchenko2003The140}, \citet[][]{Fahed2011SpectroscopyPassage}, and \citet[][]{Thomas2021The140}, as well as relative astrometric measurements from \citet[][]{Monnier2011First140} and \citet[][]{Thomas2021The140}. The dataset contains $465$, $460$, and $11$ data points for the Wolf-Rayet radial velocities, the O-star radial velocities, and the relative astrometry respectively.

We first apply the $\chi^2$ algorithm to the data and check that we recover predicted parameter distributions that are consistent with \citet[][]{Thomas2021The140}. Next, we apply our marginalised $\mathcal{GP}$ algorithm to the same data. In this case the different spectral lines are combined into one radial-velocity curve per stellar component and so we make use of only the time dimension in the GP. We report the results in Table \ref{tab:application_to_prototypical_systems_wr_140}. We find that the distributions of all the radial-velocity parameters become more uncertain, as does the semi-major axis. However, the inclination and longitude of the ascending node are fairly consistent owing to their dependence being primarily related to the astrometric fitting procedure which remains unchanged between the two algorithms. Additionally, we test the predicted parameter distributions when only the radial-velocity data are utilised. In this case we find all of the orbital elements remain ${\sim}1\sigma$ consistent and the distributions show similar variance. However, there is one exception. We find a noticeable increase in variance in the predictive distribution for the argument of periastron, $\omega_{\scalebox{0.6}{\rm{WR}}}=226.03^{+1.98}_{-1.94}$. This is somewhat expected as the astrometry data naturally helps to constrain $\omega_{\scalebox{0.6}{\rm{WR}}}$.

In the top panel of Figure \ref{fig:wr_140_gp} we display a snapshot of the conditioned GP for WR 140 for the most recent periastron passage. In the middle panel we show all of the radial-velocity data utilised in the parameter estimation. In the close-up view of periastron we observe systematics in the Wolf-Rayet's data that are reminiscent of the systematics presented in our synthetic example in Figure \ref{fig:synthetic_sb1_example_violin_1}. The median characteristic length scale of the marginalised GP kernel is $130$ days for the Wolf-Rayet's data. For the O-star, as expected, the data show only minor deviations from the orbital model, and the amplitude of the GP kernel reflects this through having a $3.4$-times smaller value relative to the Wolf-Rayet's kernel. Through applying our marginalised $\mathcal{GP}$ algorithm to these data we are able to make more accurate parameter predictions for WR 140. In this case the resulting parameter distributions, given the full complexity of the data, are more uncertain than previously predicted.

\begin{table}
    \centering
    \renewcommand{\arraystretch}{1.3}
    \begin{tabular}{l c c}
    \hline
    WR 133's orbit & \citet[][]{Richardson2021TheOrbit} & marginalised $\mathcal{GP}$ \\
    \hline
    \multicolumn{3}{c}{Elements of the system} \\
    \hline
    $P$ & $112.780^{+0.036}_{-0.036}$ & $112.763^{+0.030}_{-0.032}$ \\
    $T_0$ & $2458702.08^{+0.38}_{-0.38}$ & $2458701.90^{+0.32}_{-0.30}$ \\
    $e$ & $0.3558^{+0.0050}_{-0.0050}$ & $0.3578^{+0.0043}_{-0.0042}$ \\
    $\omega_{\scalebox{0.6}{\rm{WR}}}$ & $45.3^{+6.1}_{-6.1}$ & $47.5^{+6.5}_{-5.4}$ \\
    \hline
    \multicolumn{3}{c}{Elements of the radial-velocity orbit} \\
    \hline
    $k_{\scalebox{0.6}{\rm{WR}}}$ & $32.30^{+3.02}_{-3.02}$ & $34.04^{+4.49}_{-4.50}$ \\
    $k_{\scalebox{0.6}{\rm{O}}}$ & $14.63^{+1.51}_{-1.51}$ & $13.91^{+1.23}_{-1.30}$ \\
    $\gamma_{\scalebox{0.6}{\rm{WR}}}$ & $78.1^{+3.0}_{-3.0}$ & $79.0^{+3.4}_{-3.5}$ \\
    $\gamma_{\scalebox{0.6}{\rm{O}}}$ & $-15.09^{+0.48}_{-0.48}$ & $-14.00^{+0.88}_{-0.91}$ \\
    \hline
    \multicolumn{3}{c}{Elements of the astrometric orbit} \\
    \hline
    $a$ & $0.7863^{+0.0060}_{-0.0060}$ & $0.7838^{+0.0052}_{-0.0043}$ \\
    $i$ & $160.44^{+1.86}_{-1.86}$ & $161.15^{+1.46}_{-1.55}$ \\
    $\Omega$ & $171.5^{+6.5}_{-6.5}$ & $174.1^{+7.0}_{-5.7}$ \\
    \hline
    \end{tabular}
    \caption{The same as Table \ref{tab:application_to_prototypical_systems_wr_140} except for WR 133.}
    \label{tab:application_to_prototypical_systems_wr_133}
\end{table}

In the bottom panels of Figure \ref{fig:wr_140_gp} we display the simultaneous fit to the relative astrometric data. Given the availability of this data we can infer dynamical masses. Our results lead to the Wolf-Rayet star having a mass, $m_{\scalebox{0.6}{\rm{WR}}} = 9.90^{+1.08}_{-1.01}$, and the O-star having a mass, $m_{\scalebox{0.6}{\rm{O}}} = 27.41^{+2.50}_{-2.34}$. These masses are lower than previously estimated by \citet[][$m_{\scalebox{0.6}{\rm{WR}}} = 10.31^{+0.45}_{-0.45}$, $m_{\scalebox{0.6}{\rm{O}}} = 29.27^{+1.14}_{-1.14}$]{Thomas2021The140}, however the distributions do still have overlap.

\subsubsection{WR 133}
\label{subsubsec:wr_133}
WR 133 is an SB2 and visual system with a ${\sim}112.8$ day period at a distance of $1.86^{+0.08}_{-0.08} \kpc$ \citep[][]{GaiaCollaboration2016TheMission, GaiaCollaboration2021GaiaProperties}. The stellar components are a nitrogen-rich Wolf-Rayet star (WN5o) and an O-star (O9I). We use the orbital parameters recently calculated by \citet[][]{Richardson2021TheOrbit} as the current literature estimates.

We utilise the same data as \citet[][]{Richardson2021TheOrbit}: radial-velocity measurements from \citet[][]{Underhill1994A190918} and \citet[][]{Richardson2021TheOrbit}, as well as relative astrometric measurements also from \citet[][]{Richardson2021TheOrbit}. The dataset contains $60$ radial-velocity measurements for each stellar component, and $8$ relative astrometric measurements.

We first apply the $\chi^2$ algorithm to the data and check that we recover predicted parameter distributions that are consistent with \citet[][]{Richardson2021TheOrbit}. Next, we apply our marginalised $\mathcal{GP}$ algorithm to the same data. The different spectral lines are combined into one radial-velocity curve per stellar component and so we make use of only the time dimension in the GP. We report the results in Table \ref{tab:application_to_prototypical_systems_wr_133}. We find that the distributions of all the parameters are consistent with the literature predictions.

In the top panel of Figure \ref{fig:wr_133_gp} we display a snapshot of the conditioned GP for WR 133. We observe little to no systematics in both the Wolf-Rayet's and O-star's radial-velocity data. In the middle panel we show all of the radial-velocity data folded on the predicted period. Here we see how the deviations from the orbital model appear as white noise. As a result the median amplitude of the marginalised GP kernels for each star are at the ${\sim}3 \kmpers$ level: a barely noticeable amount given the magnitude of the semi-amplitudes.

Through applying our marginalised $\mathcal{GP}$ algorithm to these data we are able to show how, when the noise present is predominately not correlated, the estimated parameters recover the same distributions as the $\chi^2$ algorithms used throughout the literature. The marginalised $\mathcal{GP}$ is therefore safe to use in both cases of white noise and correlated noise. As for the dynamical masses of WR 133, our results lead to consistent predicted mass distributions with those of \citet[][]{Richardson2021TheOrbit}. We find the Wolf-Rayet star has a mass, $m_{\scalebox{0.6}{\rm{WR}}} = 8.9^{+3.4}_{-2.2}$, and the O-star has a mass, $m_{\scalebox{0.6}{\rm{O}}} = 21.9^{+10.1}_{-7.6}$. As noted by \citet[][]{Richardson2021TheOrbit} the large uncertainties are mainly due to the nearly face-on geometry of the system, and they improve the precision of the masses through fixing the system at the Gaia distance.

\section{Summary and conclusions}
\label{sec:summary_and_conclusions}
In this study we aimed to create a new approach for estimating the orbital parameters of emission-line binaries. Our work is summarised as follows:
\begin{enumerate}
  \item We synthesised benchmark datasets for testing the performance of different algorithms for Bayesian parameter estimation of the orbits of SB1 systems. The noise added to each dataset was designed to challenge potential models to be accurate in the presence of both white noise and problematic systematics. We have made these datasets freely available with the aim of making model validation an easy and standardised practice.
  \item We presented a novel application of Gaussian processes to model the radial-velocity systematics of emission-line binaries, named marginalised $\mathcal{GP}$. We benchmarked this algorithm, along with current standardised algorithms, on the synthetic datasets. We found that the marginalised $\mathcal{GP}$ algorithm performs significantly better than the standard algorithms when the data include complex systematics. Additionally, when the data contain no systematics the marginalised $\mathcal{GP}$ algorithm recovered a similar performance to the standardised algorithms.
  \item We applied the marginalised $\mathcal{GP}$ algorithm to four prototypical emission-line binaries. For Eta Carinae, GG Carinae, and WR 140 we found obvious systematics in the radial-velocity data, and as a result the orbital parameter distributions predicted by our marginalised $\mathcal{GP}$ algorithm are more uncertain than those claimed in the literature. For WR 133 we did not discern any systematics and our predicted distributions are consistent with those in the literature.
  \item Overall, given the validation of the marginalised $\mathcal{GP}$ algorithm on the benchmark datasets, we expect the estimated parameter distributions for Eta Carinae, GG Carinae, and WR 140 to be more accurate than those previously calculated. As a direct consequence, the dynamical masses inferred for these systems may need to be updated.
\end{enumerate}
The algorithm presented in this study represents a new baseline against which investigators should confront their models. In the future we encourage the continued development of new and ingenious algorithms to improve upon our scores on the synthetic data, and thereby help the field make increasingly better estimates of the orbits and dynamical masses of emission-line binaries. Possible performance enhancements may be found through utilising different samplers, GP kernels, mean models, parameter combinations, and other algorithms entirely. Additionally, accompanying radial-velocity data from companion stars less affected by systematics, and well-sampled astrometry is paramount in driving down parameter uncertainties when the emission-line star is heavily affected by systematics. 

\section*{Acknowledgements}
\label{sec:acknowledgements}
A great many organisations and individuals have contributed to the success of the Global Jet Watch observatories and these are listed on \url{http://www.GlobalJetWatch.net} but we
particularly thank the University of Oxford and the Australian Astronomical Observatory. We would like to thank the anonymous referee for a helpful and constructive report. We would also like to thank David Rindt for helpful discussions, and Jonathan Patterson for his support of the Oxford Physics computing cluster. We gratefully acknowledge the use of the following software: \texttt{scipy} \citep[][]{Jones2001SciPy:Python}, \texttt{numpy} \citep[][]{VanDerWalt2011TheComputation}, \texttt{pandas} \citep[][]{McKinney2010DataPython}, \texttt{george} \citep[][]{Ambikasaran2015FastProcesses}, \texttt{emcee} \citep[][]{Foreman-Mackey2013Emcee:Hammer}, and \texttt{matplotlib} \citep[][]{Hunter2007Matplotlib:Environment}. Software and benchmark datasets developed for this research are open-source and freely available at \url{https://github.com/DavoGrant/}.

\section*{Data availability}
\label{sec:data_availability}
The synthetic datasets underlying this article are available at \url{https://github.com/DavoGrant/benchmark-datasets-emission-line-binaries}. This research has made use of the data archive for Eta Carinae which is available online at \url{http://etacar.umn.edu}. The archive is supported by the University of Minnesota and the Space Telescope Science Institute under contract with NASA. This research also made use of data on GG Carinae from the Global Jet Watch. The Global Jet Watch data will be shared on reasonable request to the corresponding author. Data underlying the work on WR 140 is derived from sources in the public domain: radial-velocity measurements from \citet[][]{Marchenko2003The140}, \citet[][]{Fahed2011SpectroscopyPassage}, and \citet[][]{Thomas2021The140}, as well as relative astrometric measurements from \citet[][]{Monnier2011First140} and \citet[][]{Thomas2021The140}. Data underlying the work on WR 133 is also derived from sources in the public domain: radial-velocity measurements from \citet[][]{Underhill1994A190918} and \citet[][]{Richardson2021TheOrbit}, as well as relative astrometric measurements also from \citet[][]{Richardson2021TheOrbit}.




\bibliographystyle{mnras}
\bibliography{references} 



\appendix


\bsp	
\label{lastpage}
\end{document}